\documentclass[aps,preprint,amsmath,amssymb]{revtex4}
\usepackage{graphicx}
\usepackage{dcolumn}
\usepackage{bm}
\usepackage{amssymb}
\usepackage{amsmath}

\begin{document}


\title{Spectral Detrended Fluctuation Analysis and
Its Application to Heart Rate Variability Assessment}

\author{Luciano da Fontoura Costa$^1$} 
\email{luciano@if.sc.usp.br}

\author{Ruth Caldeira de Melo$^2$}
\author{\'Ester da Silva$^3$}
\author{Audrey Borghi-Silva$^2$}
\author{Aparecida Maria Catai$^2$}

 \affiliation{// 1 -- Instituto de F\'{\i}sica de S\~{a}o Carlos, Universidade
 de S\~{a}o Paulo, Av. Trabalhador S\~{a}o Carlense 400, Caixa Postal
 369, CEP 13560-970, S\~{a}o Carlos, S\~ao Paulo, Brazil \\
2 -- N\'ucleo de Pesquisas em Exerc\'{\i}cio F\'{\i}sico,
Laborat\'orio de Fisioterapia Cardiovascular, Universidade Federal
de S\~ao Carlos, SP, Brazil \\
3 -- Faculdade de Ci\^encias da Sa\'ude, Universidade Metodista de
Piracicaba, SP, Brazil}

\date{5th June 2005}

\begin{abstract}

Detrend fluctuation analysis (DFA) has become a choice method for
effective analysis of a broad variety of nonstationary signals. We
show in the present article that, provided the nonstationary
fluctuations occur at a large enough time scale, an alternative
approach can be obtained by using the Fourier series of the signal.
More specifically, signal reconstructions considering Fourier series
with increasing number of higher spectral components are subtracted
from the signal, while the dispersion of such a difference is
calculated.  The slope of the loglog representation of the dispersions
in terms of the time scale (reciprocal of the frequency) is calculated
and used for the characterization of the signal.  The detrend action
in this methodology is performed by the early incorporation of the low
frequency spectral components in the signal representation.  The
application of the spectral DFA to the analysis of heart rate
variability data has yielded results which are similar to those
obtained by traditional DFA.  Because of the direct relationship with
the spectral content of the analyzed signal, the spectral DFA may be
used as a complementary resource for characterization and analysis of
some types of nonstationary signals.

\end{abstract}


\maketitle

\section{Introduction}

Several natural signals are characterized by nonstationarity, i.e. the
variation of its statistical description along time.  Introduced by
Peng and collaborators~\cite{Peng:1994}, detrend fluctuation analysis
--- DFA --- has proven to be particularly effective for coping with
nonstationary signals, providing results which tend to be more robust
than previous methods such as those based on the Hurst coefficient.
One of the interesting features of DFA is its ability to reveal the
fractal nature (e.g.~\cite{Bunde_Havlin1,Bunde_Havlin2}) of
one-dimensional signals.  A particularly impacting application of DFA
has been the characterization of heart rate variability (HRV)
(e.g.~\cite{Iyengar:1996}), as estimated from sequences of interbeat
intervals (RRi).  The traditional approach to such a problem involves
normalizing the interbeat sequence so as to achieve null mean and unit
variance, integrating the resulting sequence, dividing it into boxes
of fixed size, obtaining linear fit (or higher polynomial
approximations) inside each box, and estimating the standard deviation
of the difference between the sequence and linear fit.  The absolute
value of the inclination of the loglog curve yields the $\alpha$
parameter, which has been shown to be useful for applications to
several analyses.  Note that all logarithms in this article are
natural logarithms.

Many interesting results have been obtained by DFA analysis of RRi
sequences, including the fact that the coefficient $\alpha$ provides
good sensitivity to the age of the analyzed individuals, with younger
subjects tending to present smaller values of $\alpha$
(e.g.~\cite{Iyengar:1996, Acharya:2004,
Lipsitz_circulation,Peng_Mietus:2002}).  This effect is to a great
extent a consequence of the fact that the heart rate of youngers tend
to exhibit higher small scale variability, which populates the
lefthand side of the loglog curve, therefore decreasing the value of
$\alpha$.  At the same time, complementary studies have shown that the
effect of exercise in humans tend to lead to a reduction of the value
of $\alpha_1$, namely the slope of the loglog curve considering 4 to
11 heatbeats\cite{Tulppo:2001}.  As with the aging effect, such a
reduced slope is also a consequence of the enhancement of small scale
information in the interbeat interval signals.  Additional works
dealing with the effect of exercise and physical fitness on heart
dynamics include but are not limited to~\cite{Mikko:2002,
Hautala:2003, Arja:2004, Arto:2004}.

While traditional DFA analysis has consolidated itself as an important
method for analysing heart dynamics, it is also interesting to
consider modifications of the primary approach which can lead to
eventual complementation of interpretation or improvements.  For
instance, Willson and collaborators ~\cite{Willson_1, Willson_2} have
shown that the $\alpha$ coefficient in DFA can be related to the
frequency-weighted Fourier spectrum of the original signal. The
present work reports on a modified DFA method where the piecewise
partition and fitting of the original signal is replaced by its
respective Fourier series with increasing number of coefficients.  In
addition to its inherent simplicity -- it only involves calculating
the fast Fourier transform and standard deviations, the proposed
methodology bears a direct relationship with the progressive
approximation of the signal by its Fourier series.  The potential of
the proposed DFA variant is illustrated with respect to human
interbeat interval analysis involving young and older subjects
submitted or not to regular exercises.  It is shown that the spectral
DFA leads to results which are similar to those obtained by using the
traditional DFA regarding the effects of age and physical
conditioning.

This article starts by briefly revising the traditional DFA,
describing the spectral DFA, and illustrating its application in the
characterization of interbeat interval analysis by using the
traditional and spectral DFA methods.

\section{Traditional DFA}

Typically, ECG is recorded during a fixed total period of time, and
the interbeat intervals are extracted afterwards (e.g. by considering
the signal peaks).  The interbeat intervals values are represented as
a discrete sequence $r$, such that $r(k)$ corresponds to the k-th
interbeat interval, with $k = 1, 2, \ldots, N$.  Let $\left< r
\right>$ and $\sigma_r$ stand for the average and standard deviation
of the sequence $r(k)$.  A normalized sequence $s(k)$ is obtained
through the transformation
\begin{equation}
  s(k) = \frac{r(k)- \left< r \right>}{\sigma_r}   \label{eq:normal}
\end{equation}

It can be shown that $s(k)$ has null average and unit variance.  The
sequence $s(k)$ is then divided into boxes of size $L = L_{min},
\ldots, L_{max}$ and linear interpolation is performed within each box
considering several box sizes.  The difference $d(k)$ between $s(k)$
and the obtained regression is calculated for each box.  The standard
deviation $sd(L)$ of the difference sequence $d(k)$ is estimated, and
a loglog curve is obtained for $sd(L)$ in terms of $L$.  In case the
loglog curve corresponds to a straight line, the original sequence
$r(k)$ is understood to present fractal strucutre, with the absolute
value of the inclination of the loglog curve, called $\alpha$,
providing a parameter which has shown to be particularly useful for
characterization of the heart rate variability, especially regarding
the age of the individuals~\cite{Iyengar:1996}, as well as the effect
of aerobic training~\cite{Tulppo:2001}. 

\begin{figure}
\begin{center}
\includegraphics[scale=0.5]{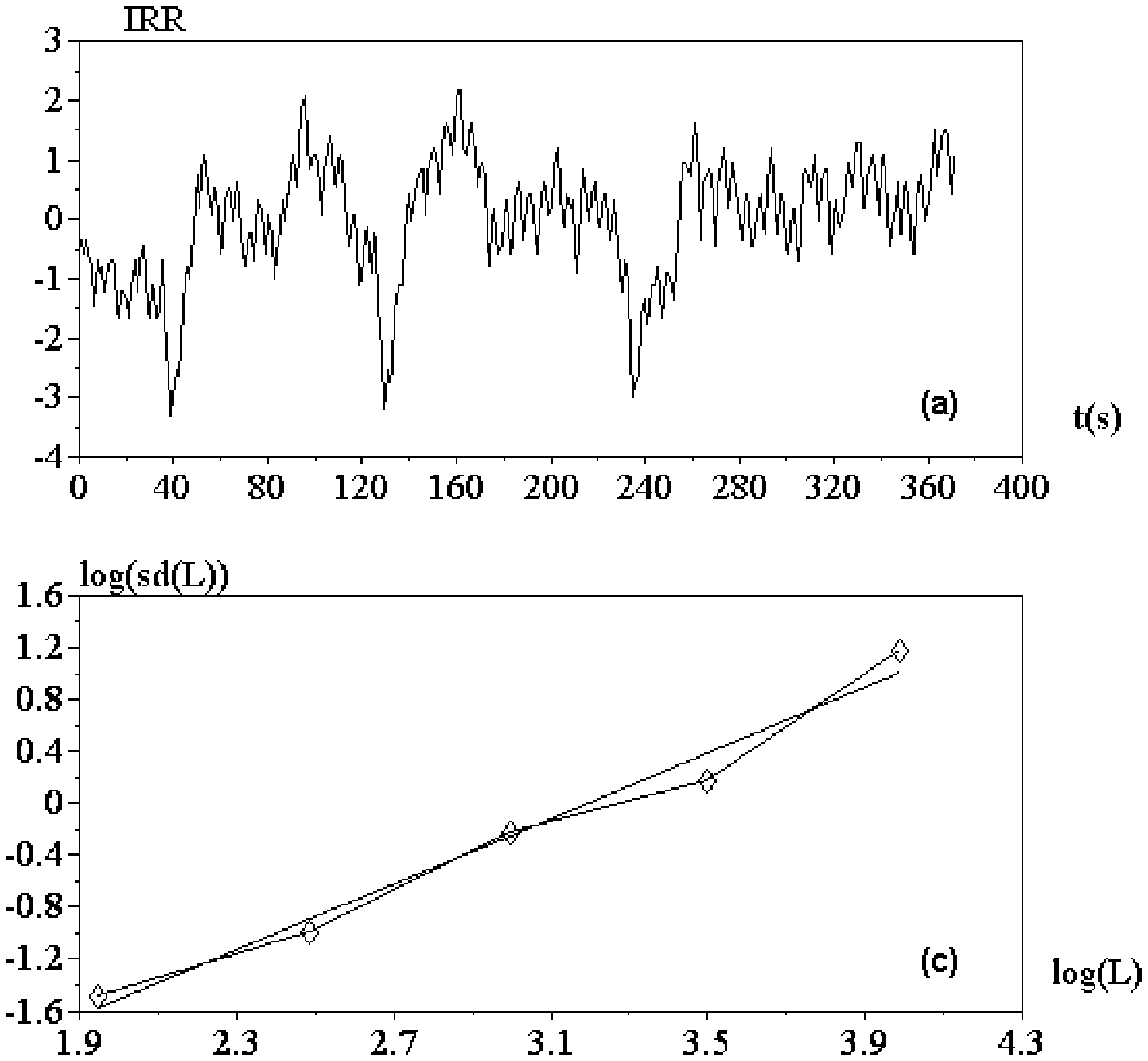}  \hspace{0.5cm}
\includegraphics[scale=0.5]{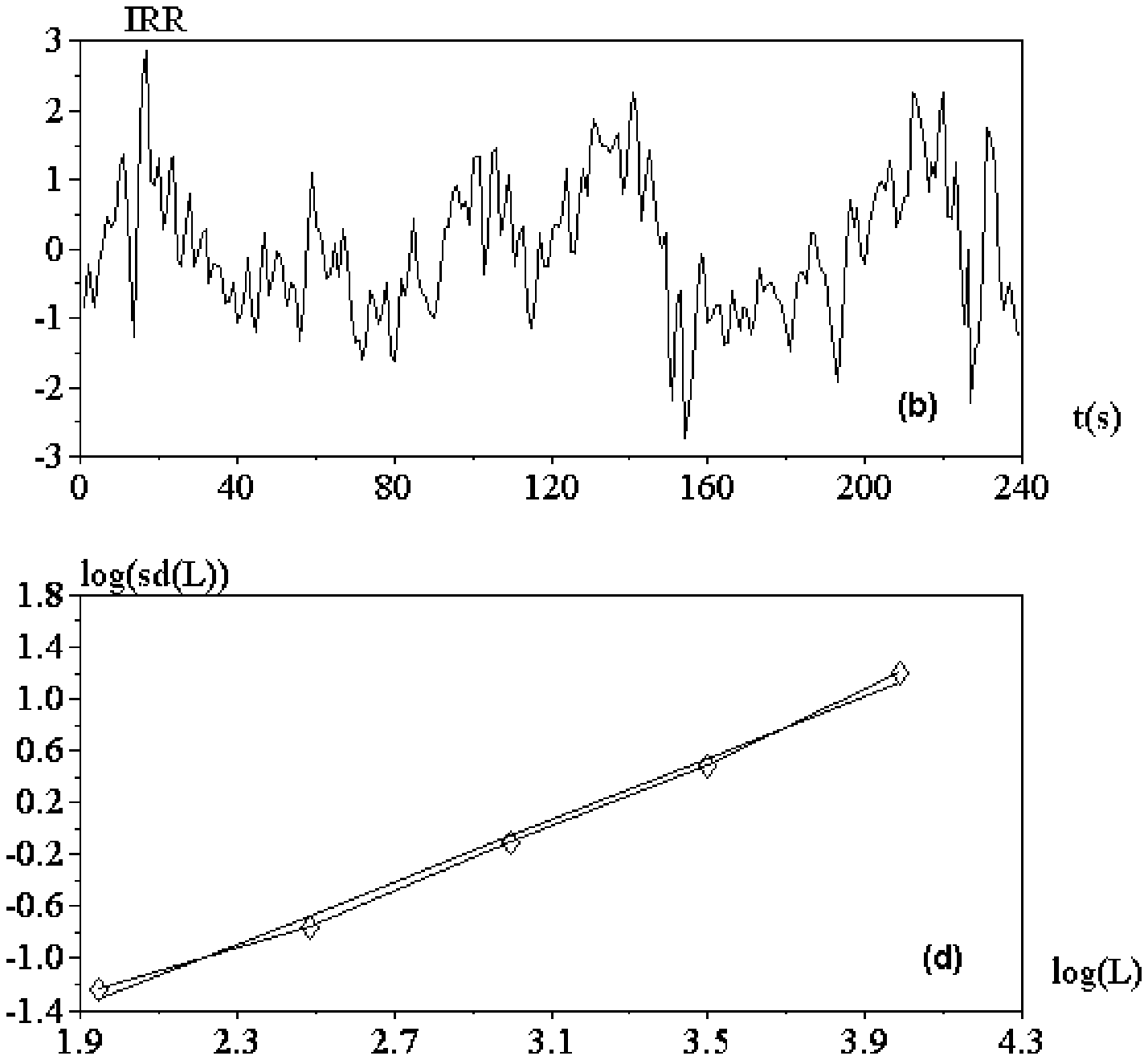}  \\

\caption{The RRi time series for a sedentary (a) and active (b) older
subjects. The respective loglog curves are shown in (c) and (d). The
values of $\alpha$ (i.e. the curve slope) obtained for these curves
were 0.98 and 0.94, respectively.}
\label{fig:DFA}

\end{center}
\end{figure}

Figure~\ref{fig:DFA} illustrates two RRi signals typical of a
sedentary (a) and active older subjects (b) as well as the respective
loglog curves (c-d), considering box sizes ranging from 4 to 11
heartbeats.

\subsection{Spectral DFA}

The Fourier series of the normalized interbeat interval signal $s(k)$
can be calculated in matrix form as

\begin{eqnarray}
  S = W s \\
  where \nonumber \\
  W= [w_{i,j}], w_{i,j} = exp \{ -  2 \pi \sqrt{-1} \: i j / N   \} \\
  i, j = 0, 1, \ldots, N-1 \nonumber
\end{eqnarray}

The reconstruction $s_m(k)$ of $s(k)$ considering the $m$ first
Fourier coefficients (i.e. $0, 1, \ldots, m-1$) can be obtained as

\begin{eqnarray}
  s_m = Q S \\
  where \nonumber \\
  Q = [q_{i,j}], \\
  q_{i,j} = 
  \begin{cases}
     exp \{  2 \pi \sqrt{-1} \: i j /N \} &  i=0, 1, \ldots, N-1; \: j= 0, 1, \ldots, m-1 \\
    0  &  i=0, 1, \ldots, N-1; \: j= m, \ldots, N-1 \nonumber
  \end{cases}  
\end{eqnarray}

Note that this equation corresponds to the Fourier series of the
signal considering m spectral components, where the quantity $u=ij$ is
proportional to the frequency.  The difference between the original
signal and its reconstruction is therefore given as $d(k) = s(k) -
s_m(k)$, whose standard deviation is henceforth expressed as $sd(m)$.

The spectral DFA proposed in this work involves calculating $sd(m)$
for several values of $m$, ploting the curve of $log(sd(m))$ in terms
of $log(1/m) = -log(m)$, and obtaining the slope $\gamma$, which has a
similar role as the parameter $\alpha$ in the traditional DFA.  Note
that as the values of $m$ are proportional to the Fourier series
frequencies, the quantity $1/m$ is proportional to the period of the
sinusoidal functions in the Fourier kernel.  Thus, $1/m$ can be
understood as a time scale parameter (analogous to the box size in the
traditional DFA) underlying the modified DFA analysis.

\begin{figure}
\begin{center}
\includegraphics[scale=0.7]{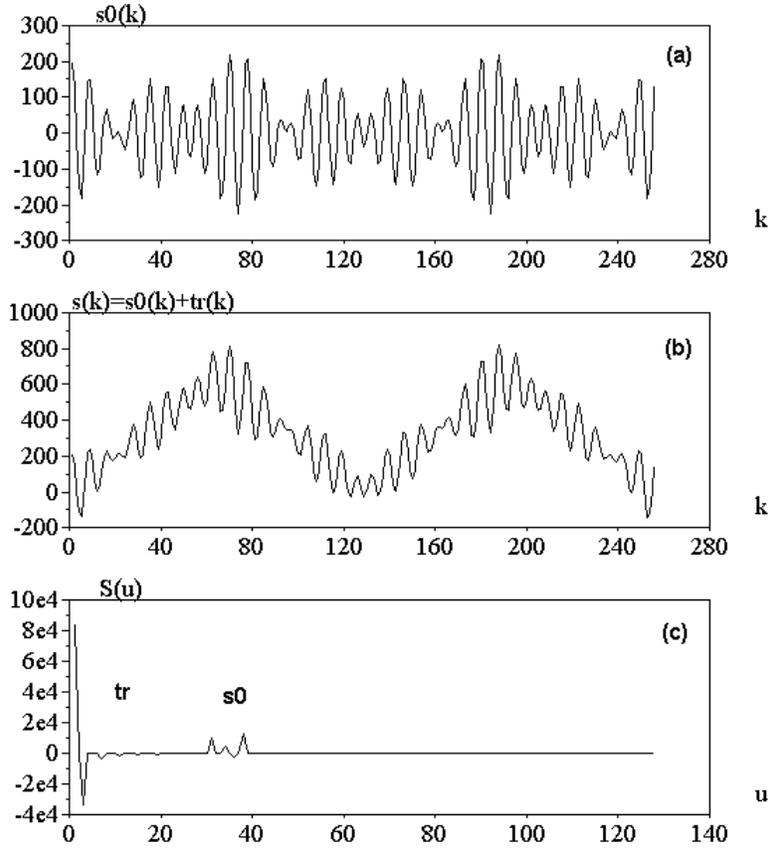} 

\caption{\label{fig:trend} An original signal of interest $s0(k)$ (a),
its contamination by a triangular function (b), and the respective
Fourier power spectrum (c).}
\end{center}
\end{figure}

The detrending effect of the above described approach is illustrated
in Figure~\ref{fig:trend}.  The original signal $s0(k)$ in (a) is
contaminated with a triangular trend function $tr(k)$, yielding the
signal $s(k)$ to be analyzed (b).  Because of the slow variation of
such a trend function, the Fourier power spectrum of $s(k)$ can be
separated into two groups, one with lower frequencies containing the
trend signal, at the left hand side in (c), and the other corresponding
to the original signal $s0$, at the righthand side in (c).  Provided
the original signal and trend contamination are reasonably
well-separated along the spectral decomposition, the trend effect can
be removed by the above difference procedure before the signal of
interest start influencing the slope $m$.  In other words, the
dispersion of the difference signal becomes independent of the trend
function.

\begin{figure}
\begin{center}
\includegraphics[scale=0.5]{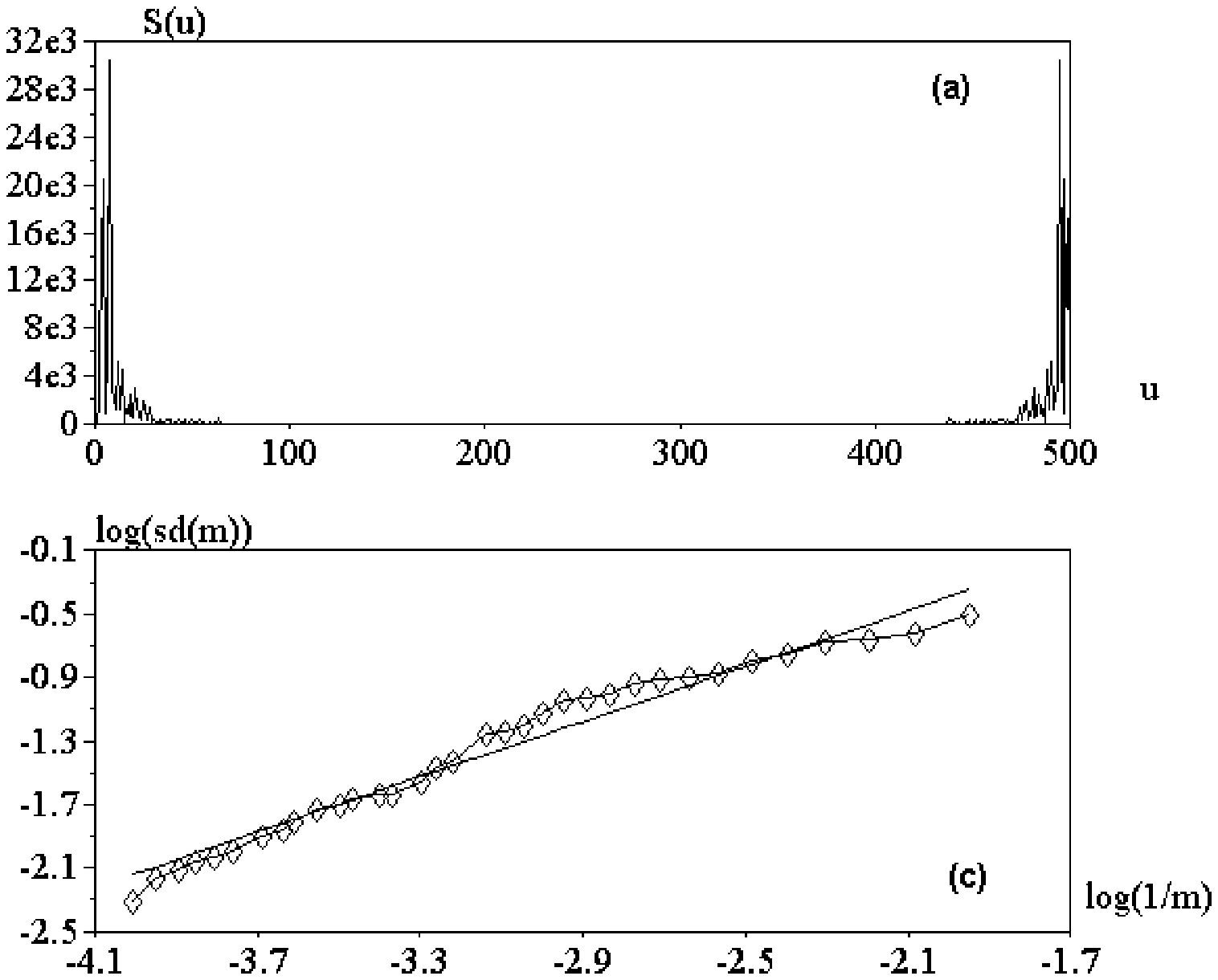} \hspace{0.5cm}
\includegraphics[scale=0.5]{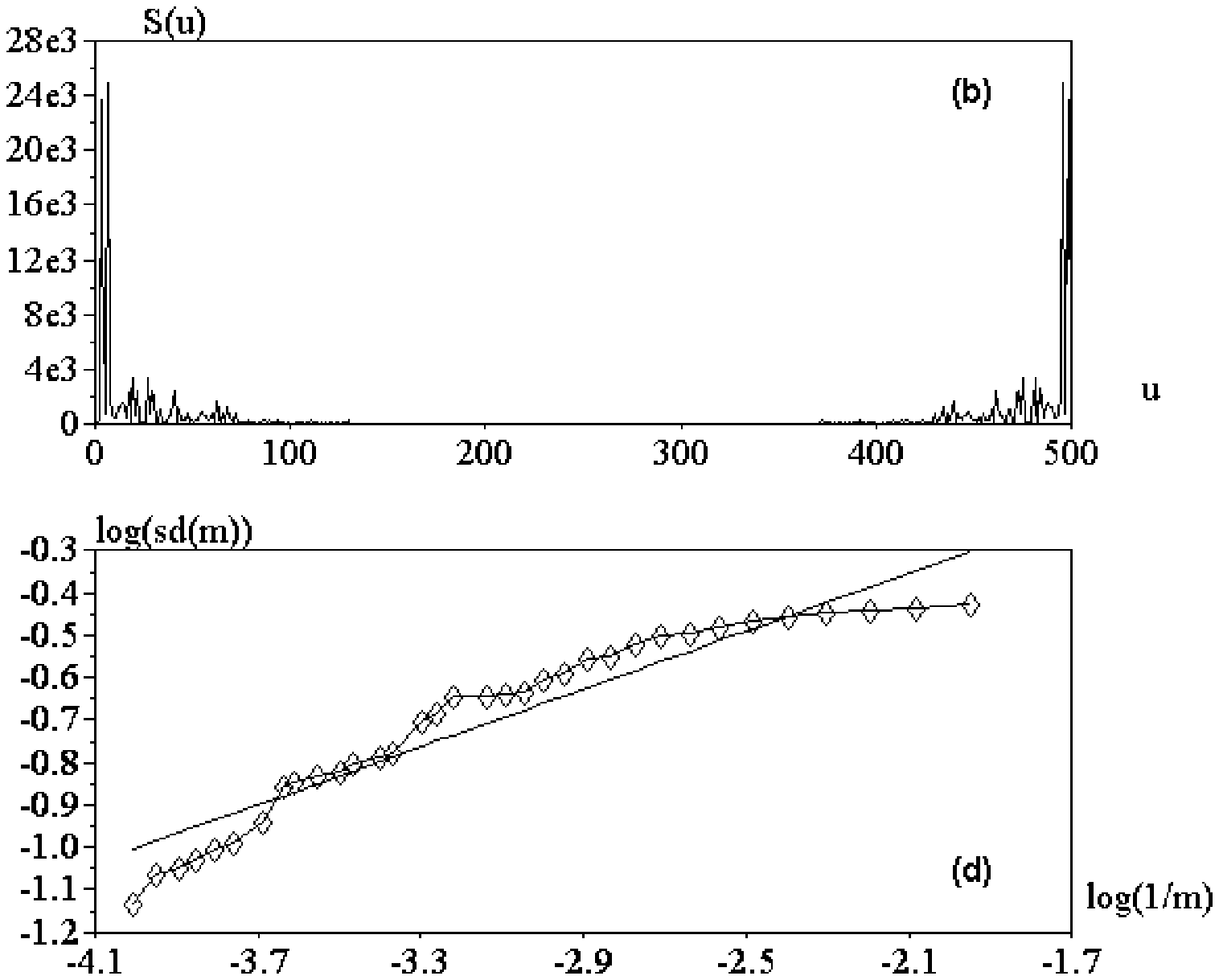}  \\

\caption{\label{fig:FDFA} The Fourier spectrum for the same sedentary
(a) and active (b) older subjects considered in Figure~\ref{fig:DFA}.
The respective loglog curves, shown in (c) and (d), have slopes 
of 0.87 and 0.34, respectively.}
\end{center}
\end{figure}

Figures~\ref{fig:FDFA}(a) and (b) illustrate the Fourier power
spectrum for the same sedentary and active older individuals in
Figure~\ref{fig:DFA}. Note that the interbeat dynamics of the active
older, whose power spectrum is shown in (b), involves a more intense
higher frequency composition.  The respective loglog curves obtained
by the spectral DFA are illustrated in (c) and (d), respectively.  It
is clear from these curves that physical exercise has as effect the
reduction of the inclination of the loglog curve, indicating that
regular physical activity tends to add small scale detail (i.e. high
frequency components) to the interbeat dynamics.  Note that the second
curve in Figure~\ref{fig:FDFA} is not well fitted by the linear
regression line.  This is a consequence of the fact that the Fourier
series approximations considered in the modified DFA methodology
contain \emph{all} coefficients ranging from $0$ to $m$, implying
faster convergence to the original interbeat signal than that which
could be obtained by using polynomial piecewise fitting.  Note the
substantially higher inclination (slope) of the curve obtained for the
sedentary subject, reflecting its lack of high frequency composition.

\section{Interbeat Interval Analysis}

In order to illustrate the performance and features of the Fourier
DFA, this method has been applied to a database of 40 interbeat
interval sequences.  

The RRi database includes measuremeents obtained at the same time of a
day from 40 male individuals divided into four classes according to
age and fitness (i.e. sedentary or active), all with good health as
confirmed by extensive clinical, physical and laboratory examinations
detailed in previous studies~\cite{Catai1}.  The individual included
in the active classes had been through regular physical activities
over a long period of time (at least 15 years in the case of the older
group).  On the day of experiments, after further medical
examinations, the subjects remained at rest in supine position during
20 minutes and then ECG was recorded for 15 minutes.  The ECG and
heart rate were acquired through a one-channel heart monitor (ECAFIX
TC500, Sp, Brazil) and processed by using an analog-digital converter
Lab.PC+ (National Instruments, Co., Austin, TX, USA), interfaced to a
personal computer.  After analog to digital conversion, the R-R
interval (ms) was calculated on a beat-to-beat basis by using a
customised software (SP, Brazil)~\cite{Catai2}.  At least 5 minutes of
ECG recording, characterized by the highest stability, were considered
for the DFA investigations in the current work.

\begin{figure}
\begin{center}
\includegraphics[scale=0.7]{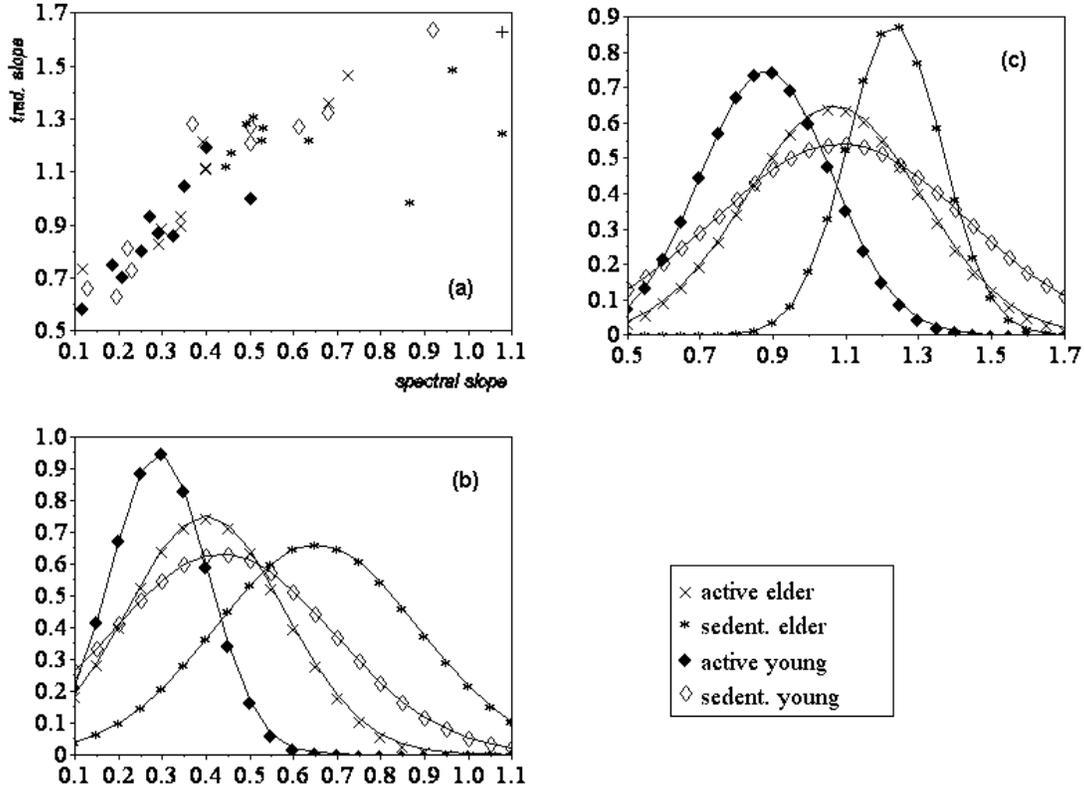} 

\caption{\label{fig:scatts} The scatterplot (a) obtained by
considering the slope of the loglog curves produced by the traditional
and spectral DFA methodologies. Density probability functions for the
four types of interbeat data obtained by normal fit of the
measurements obtained by using the traditional (b) and spectral (c)
DFA.}

\end{center}
\end{figure}

After normal transformation by using Equation~\ref{eq:normal}, each
RRi signal was processed by both traditional and spectral DFA.  The
intervals considered for calculating the loglog slopes were from 4 to
11 beats in the traditional DFA and from -4 to -2 (in log scale) in
the spectral method.  The scatterplot obtained by considering these
two measurements for each anlyzed human individual is shown in
Figure~\ref{fig:scatts} (a).  The normal density functions fitted for
each of the four classes considering each of the slopes is given in
(b) and (c).  It is clear from (a) that the results obtained by the
traditional and spectral DFA methods are highly correlated, while the
inspection of (b) and (c) indicates that similar separations between
the several classes are obtained by the two methodologies.  In both
cases, the sedentary individuals yielded loglog curve slope which is
higher than those for the active subjects.

\section{Concluding Remarks}

This article has reported on a spectral variation of the detrend
fluctuation analysis, DFA.  Provided the trend effects are slow
varying in time, the suggested methodology allows the detrend of
higher frequency components of the signal of interest.  The slope of
the loglog curve of the dispersion against time scale has been shown
to have an interpretation analogous to the $\alpha$ coefficient in the
traditional DFA.  The behavior of the spectral DFA has been
illustrated with respect to interbeat analysis considering four types
of individuals, yielding discrimination results which are closely
correlated with those obtained by traditional DFA.  Because of its
spectral nature, the new type of DFA may be of interest for some
applications in which one want to have a more direct comparison of
relationship with the spectral content of the signals.

\begin{acknowledgments}

Luciano da F. Costa thanks FAPESP (proc. 99/12765-2), CNPq
(proc. 3082231/03-1) and Human Frontier for financial support.  Ester
da Silva thanks FAPESP (proc. 01/07427-2) and Aparecida Maria Catai
thanks CNPq (proc. 478799/2003-9).

\end{acknowledgments}

\bibliographystyle{unsrt}

\bibliography{dfa}

\begin{thebibliography}{10}

\bibitem{Peng:1994}
C.-K. Peng, S.~V. Buldyrev, S.~Havlin, M.~Simons, H.~E. Stanley, and A.~L.
  Goldberger.
\newblock Mosaic organization of {DNA} nucleotides.
\newblock {\em Phys. Rev. E}, 49(2):1685--1689, 1994.

\bibitem{Bunde_Havlin2}
A.~Bunde and S.~Havlin.
\newblock {\em Fractals in Science}.
\newblock Springer, 1995.

\bibitem{Bunde_Havlin1}
A.~Bunde and S.~Havlin.
\newblock {\em Fractals and Disordered Systems}.
\newblock Springer, 1996.

\bibitem{Iyengar:1996}
N.~Iyengar, C.-K. Peng, R.~Morin, A.~L. Goldberger, and L.~A. Lipisitz.
\newblock Age-related alterationsin the fractal scaling of cardiac interbeat
  interval dynamics.
\newblock {\em Am. J. Physiol.}, 271:1078--1084, 1996.

\bibitem{Acharya:2004}
U.~R. Acharya, N.~Kannathal, S.~Ong Wai, L.~Y. Ping, and T.~Chua.
\newblock Heart rate analysis in normal subjects of various age groups.
\newblock {\em BioMedical Eng. OnLine}, 3:1--8, 2004.

\bibitem{Lipsitz_circulation}
L.~A. Lipsitz, J.~Metus, G.~B. Moody, and A.~L. Goldberger.
\newblock Spectral characteristics of hear rate variability before and during
  postural tilt.
\newblock {\em Circulation}, 81(6):1803--1810, 1990.

\bibitem{Peng_Mietus:2002}
C.-K. Peng, J.~E. Mietus, Y.~Liu, C.~Lee, J.~M. Hausdorff, H.~E. Stanley, A.~L.
  Goldberger, and L.~A. Lipsitz.
\newblock Quantifying fractal dynamics of human respiration: {A}ge and gender
  effects.
\newblock {\em Ann. Biom. Eng.}, 30:683--692, 2002.

\bibitem{Tulppo:2001}
M.~P. Tulppo, R.~L. Hughson, T.~H. Makikallio, K.~E.~J. Airaksinin,
  T.~Seppanen, and H.~V. Huikuri.
\newblock Effects of exercise and passive ehad-up tilt on fractal and
  complexity properties of heart rate dynamics.
\newblock {\em Am. J. Physiol. Heart Circ. Physiol.}, 280:1081--1087, 2001.

\bibitem{Arto:2004}
A.~J. Hautala, T.~H. Makikallio, A.~Kiviniemi, R.~T. Laukkanen, S.~Nissila,
  H.~V. Huikuri, and T.~P. Tulppo.
\newblock Heart rate dynamics after controlled training followed by a
  home-based exercise probram.
\newblock {\em Eur. J. Appl. Physiol}, 92:289--297, 2004.

\bibitem{Hautala:2003}
A.~J. Hautala, T.~H. Makikallio, T.~Seppanen, H.~V. Huikuri, and M.~P. Tulppo.
\newblock Short-term correlation properties of r-r interval dynamics at
  different exercise intensity levels.
\newblock {\em Clin. Physiol. FInct. Imaging}, 23:215--223, 2003.

\bibitem{Mikko:2002}
M.~P. Tulppo, A.~J. Hautala, T.~H. Makikallio, T.~T. Laukkanen, S.~Nissila,
  R.~L. Hughson, and H.~V. Huikuri.
\newblock Effects of aerobic training on heart rate dynamics in sedentary
  subjects.
\newblock {\em J. Appl. Physiol.}, 95:364--372, 2003.

\bibitem{Arja:2004}
A.~L.~T. Uusitalo, T.~Laitinin, S.~B. Vaisane, E.~Lansimies, and R.~Rauramaa.
\newblock Physical training and heart rate and blood pressure variability: a
  5-yr randomized trial.
\newblock {\em Am. J. Physiol. Hear Circ. Physiol.}, 286:1821--1826, 2004.

\bibitem{Willson_2}
K.~Willson and D.~P. Francis.
\newblock A direct analytical demonstration of the essential equivalente of
  detrended fluctuation analysis and spectral analysis of rr interval
  variability.
\newblock {\em Physiol. Meas.}, 24:N1--N7, 2003.

\bibitem{Willson_1}
K.~Willson, D.~P. Francis, R.~Wensel, A.~J.~S. Coats, and K.~H. Parker.
\newblock Relationship between detrended fluctuation analysis and spectral
  analysis of heart-rate variability.
\newblock {\em Physiol. Meas.}, 23:385--401, 2002.

\bibitem{Catai1}
R.~C. Melo, M.~D.~B. Santos, E.~Silva, R.~J. Quint\'erio, M.~A. Moreno, M.~S.
  Reis, I.~A. Verzola, L.~Oliveira, L.~E.~B. Martins, L.~Gallo-Junior, and
  A.~M. Catai.
\newblock {\em Braz. J. Med. Biol. Res.}, 2005.
\newblock In press.

\bibitem{Catai2}
E.~Silva, A.~M. Catai, L.~C. Trevelin, J.~O.~Guimar\ aes, L.~P. Silva-Junior,
  L.~M.~P. Silva, L.~Oliveira, L.~A. Milan, L.~E.~B. Martins, and
  L.~Gallo-Junior.
\newblock {\em Phys. Med. Biol.}, 409:39a, 1994.
\newblock Abstract.

\end{thebibliography}

\end{document}